\documentclass[onecolumn]{emulateapj}
\usepackage{amsfonts}
\usepackage{amsmath}
\usepackage{amssymb}
\usepackage{wrapfig}
\usepackage{graphicx}
\usepackage{bibentry,natbib}
\usepackage{wasysym}
\usepackage{latexsym}
\usepackage{subfigure}

\newcommand{\et}{{\it et al}.}

\newcommand{\half}{{\textstyle{1\over2}}}

\newcommand{\be}{\begin{equation}}
\newcommand{\ee}{\end{equation}}
\newcommand{\bea}{\begin{eqnarray}}
\newcommand{\eea}{\end{eqnarray}}

\def\curl{{\rm curl}\,}

\slugcomment{to be submitted to Astrophys. J.}

\shorttitle{Generic model for magnetic explosions}
\shortauthors{Melrose}

\begin{document}

\title{Generic model for magnetic explosions applied to solar flares}

\author{D. B.  Melrose}

\affil{Sydney Institute for Astronomy, School of Physics, The University of Sydney, NSW 2006, Australia}

\begin{abstract}
An accepted model for magnetospheric substorms is proposed as the basis for a generic model for magnetic explosions, and is applied to solar flares. The model involves widely separated energy-release and particle-acceleration regions, with energy transported Alfv\'enically between them. On a global scale, these regions are coupled by a large-scale current that is set up during the explosion by redirection of pre-existing current associated with the stored magnetic energy. The explosion-related current is driven by an electromotive force (EMF) due to the changing magnetic flux enclosed by this current. The current path and the EMF are identified for an idealized quadrupolar model for a flare.
\end{abstract}



\section{Introduction}
\label{section:introduction}

Recent dramatic improvements in observational resolutions of the active Sun have led to improved understanding and modeling of detailed phenomena in solar flares \citep{BG10,Hetal11,Ketal11,Whetal11}. However, this emphasis on details has left unresolved several long-standing problems relating to the global electrodynamics and energetics of flares. We do not have an accepted global model for a flare that allows a semi-quantitative estimate of the power and energy in terms of large-scale observable parameters.  A global model for solar flares is needed not only for the solar application, but also as the basis for a generic model for other magnetic explosions, such as flares on other stars \citep{BG10} and magnetar outbursts \citep{Ly06}.  For any form of magnetic energy release, integration of the source term (${\bf J}\cdot{\bf E}$) for electromagnetic energy over the volume in which the energy is stored implies that the power released can be identified as $I\Phi$. A global description of a magnetic explosion requires some form of circuit model in which an electromotive force (EMF), $\Phi$, drives a current, $I$, around a circuit. Despite interest in circuit models in the 1970s and 1980s \citep{C78,S82}, we have no acceptable model (a) for the circuit that describes how the stored energy is released, (b) for the current (intrinsically) associated with the energy release, nor (c) for the EMF that drives this current. Another long-standing ``number problem'' \citep{Betal09} is that the X-ray data imply energetic ($\varepsilon\approx10^5\,$eV) electrons precipitating at a rate ${\dot N}$ such that $e{\dot N}$ exceeds any plausible current by a very large factor, and such that the total number of electrons accelerated greatly exceeds the number initially in the flaring coronal flux tubes. A further long-standing problem is that any estimate of $\Phi$ based on the changing magnetic field implies a value, of order $10^{10}\,$V \citep{S33,C78}, that greatly exceeds $\varepsilon/e\approx10^5\,$V, and for which there is no direct evidence.  

There is a long history of relating models for solar flares and models for magnetospheric substorms, which are a terrestrial form of magnetic explosion. For example, the widely favored CSHKP model \citep{FM86} for flares involves a current sheet in an outflowing plasma that is analogous to the current sheet in the Earth's magnetotail. A major difference between flares and substorms, from a global viewpoint, is that the stored magnetic energy in a flare must be associated with currents that flow through the photosphere and close below it, whereas the stored magnetic energy in a substorm is associated with currents confined to the magnetotail. From an energetic viewpoint, a global model for a flare must involve a reconfiguration of the coronal paths of currents that flow through the photosphere \citep{M97,HMH98}, rather than a change in the magnitude of these currents. This essential ingredient is not compatible with a stand-alone CSHKP model. It has been argued \citep{Uetal99} that the CSHKP model should be regarded as one part of a global model for a flare, rather than as a global model on its own.

\begin{figure} [t]
\centerline{
\includegraphics[scale=0.7]{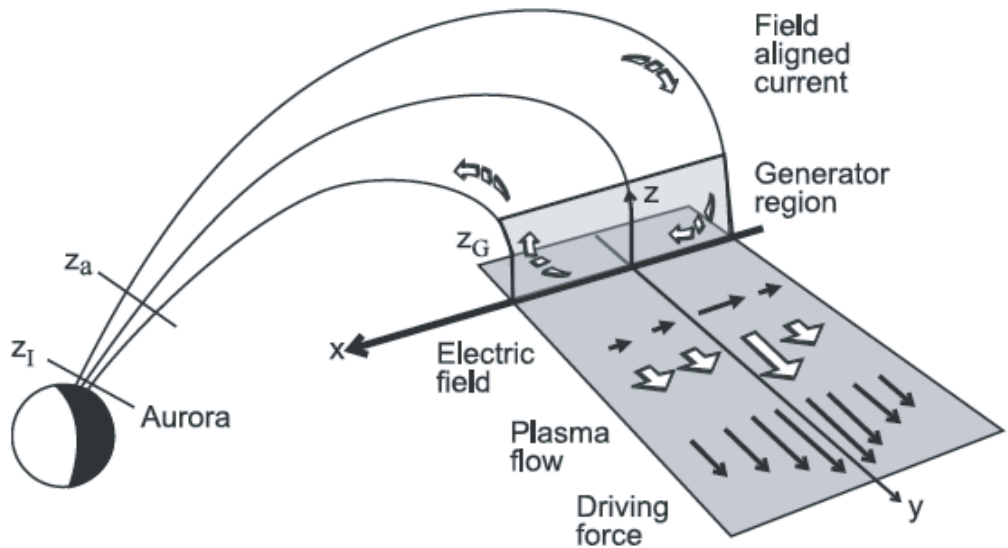}
}
\caption{Model for a substorm: the driving force (a pressure gradient) is indicated in the $x$-$y$ plane at $z=z_G$ in the magnetotail; the cross-tail current is redirected along field lines, closing in the ionosphere (labeled Aurora) at $z=z_I$; the dissipation occurs in a separate region, $z=z_a$, between these two regions. [After \cite{R02}]}
\label{fig:generator}
\end{figure}

In a magnetic explosion, magnetic energy builds up in a diffuse plasma over a relatively long time, and is released on a much shorter (explosive) time. A substantial part of the released energy appears in fast electrons (first-phase electrons in a flare and auroral inverted-V electrons in a substorm). In formulating a generic model for magnetic explosion, it is important that the model account both for the features common to flares and substorms, and for essential differences between flares and substorms. There is a widely-accepted global model for substorms \citep{McPetal73,TB97}, illustrated in Fig.~\ref{fig:generator}; some features of this model should to be included in a generic model (and applied to flares), while other features are specific to substorms.  Three specific ingredients are assumed here to be generic. First, the primary energy release site is far removed from the region where the the acceleration of electrons occurs. Second, a new current system is set up during the explosion, and this current connects the energy-release and acceleration regions. Third, the energy transport between the energy-release and acceleration regions is Alfv\'enic, involving the parallel current and a cross-${\bf B}$ electric field.  One feature that is favored for substorms but cannot apply to flares concerns the driving force: this attributed to a pressure gradient in the generator region in Fig.~\ref{fig:generator}, but pressure gradients in the corona are inadequate to drive a flare, and the driver must be an electromagnetic (Maxwell) stress. Conversely, the generic model suggests that a more plausible driver for a substorm is a Maxwell stress analogous to the the driver for a flare.

In the generic model proposed here, emphasis is placed on an essential aspect of the electrodynamics that is obscured in most models for solar flares: the intrinsic time dependence. Flares are obviously intrinsically time-dependent phenomena, implying an intrinsically time-dependent magnetic field. Nevertheless, most models for flares are steady-state: models for magnetic reconnection, electron acceleration, etc., have the intrinsic time-dependence replaced by proxies involving boundary conditions that allow inflow or outflow. Any steady-state model has $\partial{\bf B}/\partial t=0$, so that there is no inductive electric field (one with $\curl{\bf E}\ne0$) and no displacement current ($\partial{\bf E}/\partial t=0$); the electric field in the model is necessarily an electrostatic or potential  field. The physical differences between an inductive field and a potential field are ignored in most models. For some purposes,  the steady-state models are effective proxies. However, this is not the case for other purposes, notably for electron acceleration. Although a variety of alternative acceleration mechanisms remain under consideration, the seemingly simplest and most plausible mechanism, acceleration by a parallel electric field, $E_\parallel$, is so poorly understood \citep{M09,BG10} that it is not always considered as a viable mechanism for flares \citep{Zetal11}, whereas it is the accepted mechanism for the acceleration of auroral electrons in a substorm. Inclusion of explicit time-dependence is essential for understanding acceleration by $E_\parallel$ \citep{SL06}.

The generic model for magnetic explosions proposed here has one additional notable feature. Besides being based on substorms and including time-dependence explicitly, it involves an explicit separation into multiple scales. Three scales are introduced: global, macro and micro. The global scale is at least as large as an active region, and is the scale of the circuit around which the flare-associated current flows. The macro scale includes the energy-release region and the acceleration region, and the energy transport between them. The collisionless plasma processes that are essential for magnetic reconnection (in the energy-release region) and for $E_\parallel$ (in the acceleration region) occur on the micro scale, related to the Debye length and other fundamental lengths in the plasma. Such collisionless processes are highly localized and transient, and a statistically large number of them is required to have macroscopic consequences. In this paper, emphasis is placed on the global scale. Detailed models for some of the processes on the macro scale are discussed in an accompanying paper (Melrose 2012), referred to as paper~2.

The generic model for a magnetic explosion is summarized in \S\ref{section:generic}.  Application of the model to solar flares, involving identification of the current and the current path, is discussed in \S\ref{section:current}, and the EMF is discussed in  \S\ref{section:EMF}.  Comparison with the application to substorms is discussed in \S\ref{section:substorms}. The results are discussed in \S\ref{section:discussion}.

\section{Generic model for magnetic explosions}
\label{section:generic}

In this section a widely accepted model for magnetospheric substorms is reformulated as a generic model for magnetic explosions in space and astrophysical plasmas.

\subsection{Generic model}

The generic model combines several ideas: nonlocal energy release, a multi-scale approach, intrinsic time-dependence, and globally controlled dissipation. The generic model contains the following ingredients:
\begin{description}
\item[1). {\it Driver\/}:] The temporally changing magnetic field produces an EMF, $\Phi$.
\item[2). {\it Energy-release region\/}:] The EMF drives a current, $I$, across magnetic field lines in an energy-release  region and around a (new) global-scale circuit.
\item[3). {\it Energy transport\/}:] The magnetic energy released is transported away from the energy-release region as an Alfv\'enic  flux.
\item[4). {\it Current closure\/}:] The (newly directed) current flows along field lines to a remote closure region, and back to the energy-release region along neighboring field lines.
\item[5). {\it Acceleration region\/}:] The current flows through an acceleration region, where $\Phi$ partly localizes along  field lines, leading to electron acceleration by $E_\parallel$ and dissipation of the released magnetic energy.
\item[6). {\it Globally controlled dissipation\/}:] Collisionless dissipation processes on a micro scale combine to give an effective dissipation whose rate is governed by the global requirements. 
\end{description}

\subsection{Remarks on the generic model}
Several remarks on these six ingredients in the generic model are appropriate. 

1). The words ``driver'' and ``trigger'' are used with specific meanings in the context of magnetic explosions. There are two different drivers. One applies to the build-up phase, is different for substorms and flares, and is not directly relevant to the explosion. The effect of this driver is to store magnetic energy in a metastable configuration on a relatively long time scale. The trigger for a magnetic explosion initiates the (rapid) change in configuration that results in release of the stored energy; the trigger is not clearly identified in any context. The primary magnetic energy release requires magnetic reconnection, and the ``driver'' referred to here drives inflow into a reconnection region. The only possibilities for this driver are a pressure gradient \citep{S78,H07} and the Maxwell stress (the ${\bf J}\times{\bf B}$ force in MHD). The assumption that a pressure gradient is the driver is questionable for a substorm, and unacceptable for a flare. When the intrinsic time dependence is taken into account, the Maxwell stress becomes a natural driver in both cases, as discussed in detail in paper~2.  

The name ``generator region'' is used in the magnetospheric literature for the energy-release region, and this name is regarded as misleading in the present context. In both substorms and flares, pre-existing currents are redirected, rather than an intrinsically new current being generated. 

The energy release involves magnetic reconnection. However, an essential aspect that is not addressed in most detailed models for solar flares is that the reconnecting field is twisted, and carries a current that is approximately force-free. It is the redirection of this current on the global scale that releases the stored magnetic energy \citep{M97}. Some detailed calculations for reconnection are available \citep{DAN97,L06,LGL09}. Here it is simply assumed that magnetic flux and current are transferred together during reconnection.

2). Energy flow is determined by the Poynting vector, with ${\bf E}=-{\bf u}\times{\bf B}$ in ideal MHD:
\be
{\bf S}={{\bf E}\times{\bf B}\over\mu_0}
={|{\bf B}|^2\over\mu_0}\,{\bf u}-{{\bf u}\cdot{\bf B}\over\mu_0}\,{\bf B}.
\label{Poynting}
\ee
The inflow of energy into the energy-release region may be described by the first term on the right hand side of (\ref{Poynting}), with ${\bf u}\cdot{\bf B}=0$;  the outflow may be described by the second term, referred to here as an Alfv\'enic flux. The argument for Alfv\'enic energy transport has been made in connection with several flare models \citep{ES82,H94,FH08,KHB11}. In other flare models the energy outflow is assumed to consist of energetic particles and/or mass motions. 

3). Alfv\'enic energy transport involves a parallel current, a cross-${\bf B}$ electric field, and a vortex motion, which can be interpreted as a propagating magnetic twist. Processes whereby the magnetic energy released by reconnection in a flare is converted into an Alfv\'enic flux were discussed by \cite{FH08}. An idealized cylindrical model is presented in paper~2.  

4). In a substorm,  parallel currents flow up and down along flux tubes, which converge just above the ionosphere, where the current closes across field lines. This closure is usually attributed to the Pedersen conductivity, which can be described by an effective resistance,  $R_{\rm P}$, determined by the inverse of the height-integrated conductivity. In a flare, the current may close in a similar manner in a partially ionized region of the chromosphere \citep{MK89}. Alternatively, the current may close dynamically, driving an Alfv\'en wave into the denser regions of the atmosphere where the Alfv\'enic propagation time is long compared with the time scale of the explosion, due to the rapid increase of  density with depth through the chromosphere \citep{WM94}. The effective resistance of the closure region can be interpreted as $R_{\rm P}$ or the Alfv\'enic impedance, $R_A=\mu_0v_A$: Alfv\'enic energy propagation dominates for $R_A\gg R_P$, and resistive dissipation of the energy dominates for $R_P\gg R_A$. In the model assumed here, the impedance of the current-closure region is smaller than other relevant impedances, and the energy dissipation associated with this current closure is negligible. 

5). Electrons are assumed to be accelerated by $E_\parallel$, which develops by $\Phi$ becoming partly localized along the field lines. This acceleration process is inadequately understood, as discussed in paper~2. The energy released goes into bulk energization of electrons in compact flares, which are non-eruptive. In an eruptive flare, much of the energy released goes into kinetic energy of mass motion of a coronal mass ejection (CME).  The acceleration of the CME requires a modification of the generic model that is not discussed here.

6). Collisionless dissipation involves a statistically large number of  localized and transient instabilities. This allows a major simplification: one can assume that the statistical distribution adjusts to meet the global requirement that drives the dissipation. In a driven system, the dissipation is maximized when the dissipation corresponds to an effective resistance, $R_{\rm eff}$ of order $R_A$, when the energy is absorbed,  rather than being transmitted (for $R_{\rm eff}\ll R_A$) or reflected (for $R_{\rm eff}\gg R_A$). The effective dissipation associated with the conversion to an Alfv\'enic flux in the energy-release region and with the actual dissipation in the acceleration region are assumed to correspond to $R_{\rm eff}\approx R_A$.

\section{Flare-associated current}
\label{section:current}

In the application of the generic model to solar flares, unlike that to substorms (\S\ref{section:substorms}), the essential features of the model do not follow directly from observations. On the global scale, the essential features that need to be identified are the current and the EMF. The current is discussed in this section and the EMF in \S\ref{section:EMF}.

\subsection{Release of magnetic free energy}

Prior to a flare, magnetic energy is stored in a complex magnetic structure overlying an active region. Newly emerging flux tubes are twisted \citep{Letal96}, implying that they carry currents. As the magnetic flux builds up, the current builds up correspondingly.  A twisted flux loop has a current flowing from one footpoint to the other and closing at the based of the convection zone \citep{MF89}. Magnetic energy in the corona may be attributed to three different classes of current: (a)~currents that are confined to the solar interior, (b)~currents that are confined to the corona, and (c)~currents that link the interior and the corona. Currents of class~(a) produce a ``potential'' component of the coronal magnetic field that is unchanged during a flare. The energy stored in currents of class~(b) is too small to account for the energy released during a flare; specifically, the assumed closure in the corona implies ${\bf J}\times{\bf B}$ in the closure region, and this force must be balanced by a mechanical force (a pressure gradient or gravity), which severely restricts the magnitude of ${\bf J}\times{\bf B}$. Although these currents may be important during a flare, when the ${\bf J}\times{\bf B}$ be balanced the acceleration of plasma flows, they cannot be important energetically on the global scale. Only currents of class~(c), which flow along field lines in the corona, and close near the base of the convection zone \citep{MF89} are relevant to the global energy storage. The current passing through the photosphere/chromosphere can be determined from vector magnetograms, which imply currents in flaring flux tubes $\gtrsim 10^{11}\,$A, and under extreme conditions $\approx10^{13}\,$A \citep{MS68}.

The magnetic free energy is associated with the resulting current system, with all (energetically) important currents flowing through the photosphere. On a global scale, the free energy can be identified as $\half LI^2$, where $L=\mu_0\ell$ is an effective inductance, and $\ell$ has the dimensions of a length. The net current cannot change at the footpoints during a flare \citep{M95,M97}, implying that $I$ does not change during a flare. Except for regions around magnetic nulls, where reconnection occurs, these large-scale coronal currents flow along field lines, and a change in the current configuration must be associated with a change in the magnetic configuration. 

The trigger for a flare has not been identified explicitly. The newly emerging flux builds up magnetic stresses in the corona over a time scale of days, and the reconnection must allow a reconfiguration of the magnetic field and current on a large scale to account for the magnetic energy release. In an emerging-flux model, the trigger presumably involves the emerging flux tube being forced into contact with an overlying flux tube \citep{Netal97}.

In summary, the release of magnetic free energy in a flare requires that magnetic reconnection in the corona leads to a magnetic reconfiguration that reduces the effective length, $\ell=L/\mu_0$, of the current-carrying flux tubes.

\begin{figure}[t]
\centerline{
\includegraphics[scale=0.7]{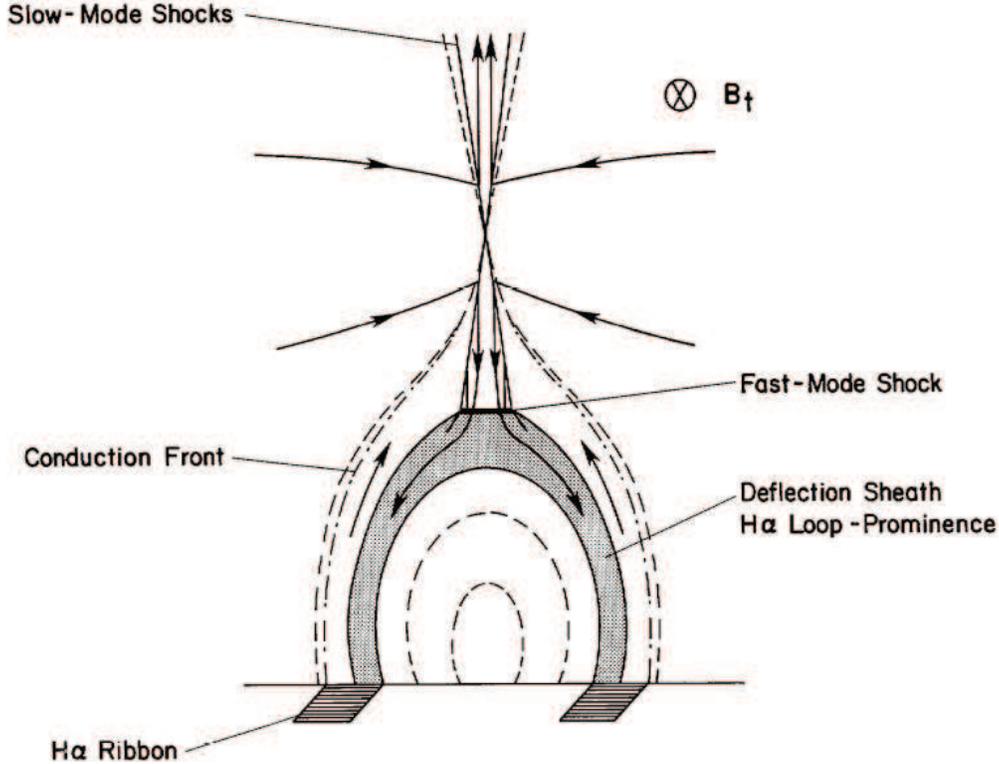}
}
\caption{The standard CSHKP model for a flare as illustrated by \cite{FM86}. In the interpretation assumed here, the reconnecting flux tube carries a current, and the energy release is due to the shortening of the current path as a reconnected flux tube moves closer to the photosphere. } 
\label{fig:CSHKP}
\end{figure}

\subsection{Flare models}

Although widely favored, the CSHKP model (Fig.~\ref{fig:CSHKP}) is inadequate as a global model for a flare. In order to discuss the global energetics, one needs a model that relates the currents in the corona to the currents passing through the photosphere, and the global current is not included in a conventional CSHKP model. This deficiency in the CSHKP model was one of the motivations for developing a quadrupolar model.

A quadrupolar model is the simplest example of a magnetic (and current) configuration that allows magnetic energy release without changing the current or magnetic flux at the photospheric boundary  \citep{M97}. Various quadrupolar models have been proposed  \citep{U96,Netal97,Uetal99,Aetal99,A04}. One version is an idealized form of an emerging flux model, in which a rising magnetic loop encounters an overlying magnetic loop. As in Fig.~\ref{fig:quadrupolar}, let the two loops be labeled 1 and 2 with footpoints $1_\pm$ and $2_\pm$, respectively. Reconnection is assumed to occur at the point of contact, $C$ say,  forming two new flux loops connecting $1_+$ to $2_-$ and $2_+$ to $1_-$ respectively. Numerical models for reconnection of two current-carrying flux loops \citep{LL06} have recently been applied to a quadrupolar configuration \citep{Tetal11}, showing how the two new flux tubes are formed. An analytic model allows the magnetic energy released to be estimated \citep{HMH98} in terms of the magnetic reconfiguration reducing the effective length of the current path. Such models describe the initial and final states, with the energy released being identified as the difference in the stored magnetic energy between these two states. The current during the flare, indicated schematically in Fig.~\ref{fig:quadrupolar}b, is discussed separately below.

A more realistic configuration for many flares involves an arcade of current-carrying magnetic loops. A shearing motion that displaces the footpoints in opposite directions on either side of the arcade increases the length of the current paths, storing magnetic energy.  Magnetic reconnection between neighboring flux tubes, which pairwise form a quadrupolar-type configuration, can lead to a net shortening of the current path, and release of the stored energy.

The shortening of the current path has potentially observable consequences. While the difficulties in identifying the change in magnetic configuration from pre-flare to post-flare are well known, vector magnetogram data suggest an increase in the transverse component of the photospheric field during a flare \citep{Wetal11}. This is consistent with the shortening of the current path: qualitatively, a shortening of the net path requires a reduction in the angle at which the magnetic field emerges through the photosphere, implying an increase in the tangential component of the field.

\subsection{Current path during a flare}

The new current path set up during a flare must have the effect of converting the initial current configuration into the final current configuration. A current that leads to this result is found by subtracting the initial current from the final current. 

\begin{figure}[t]
\centerline{
\includegraphics[scale=0.7]{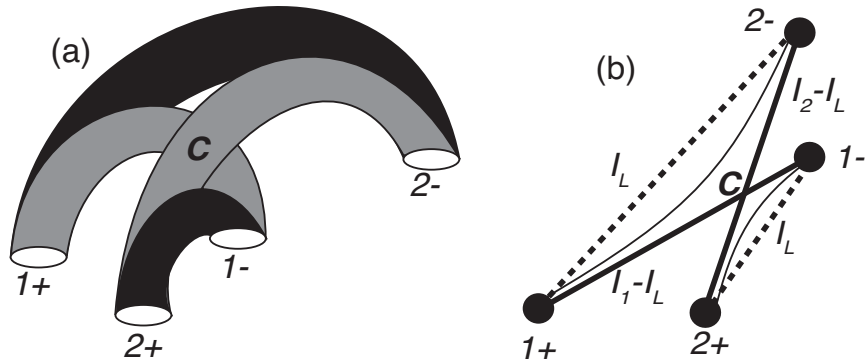}
}
\caption{Idealized quadrupolar model for a solar flare: (a) initial loops in gray, final loops in black; (b) projection onto the solar surface with the initial loops shown as thick solid lines, the final loops as dashed lines, and reconnected flux loops moving from one to the other, shown by thin lines. The EMF is assumed to drive current around the two triangles in (b), requiring current closure across field lines at each of the four footpoints. [After \cite{M97}]}
\label{fig:quadrupolar}
\end{figure}

For the quadrupolar model illustrated in Fig.~\ref{fig:quadrupolar}b, the difference between the final and initial current paths separates into two paths. Let the point where the two initial current paths (thick solid lines) reconnect at the point of contact, $C$, be identified as the energy-release region. These two paths (projected onto the solar surface) are the triangles $1_+\,2_-\,C\,1_+$ and $2_+\,1_-\,C\,2_+$. Let the initial currents between $1+,1-$ and $2+,2-$ be $I_1$ and $I_2$, respectively, and the final currents between these footpoints be $I'_1$ and $I'_2$, respectively. Conservation of current at each footpoint requires $I'_1=I_1-I_L$ and $I'_2=I_2-I_L$, with $I_L$ the final current between $1+,2-$ and $2+,1-$. The flare-associated current in both triangles is $I_L$. The current $I_L$ closes across field lines at each of the four footpoints and at $C$. Thus, there are up and down flare-associated currents at all the four footpoints where the initial and final loops meet at the photosphere/chromosphere. Downward electron acceleration, producing hard X-rays, is expected with the upward current at each of the four footpoints, by analogy with the auroral regions in substorms. 

 \subsection{Comparison with circuit models}
 
 Although the quadrupolar model involves global parameters $I$ and $\Phi$, it is not a circuit model in the conventional sense. In a circuit model \citep{C78,S82}, the current path is specified,  and the EMF and the power are associated with a photospheric dynamo. In the model illustrated in Fig.~\ref{fig:quadrupolar}, the currents flow along field lines, except at the point $C$ and the closure regions at the footpoints. During the flare, the current configuration changes due to transfer of current-carrying flux tubes from the initial to the final flux loops; the light solid line in Fig.~\ref{fig:quadrupolar}b indicates an intermediate location of a flux tube in motion. The intrinsic time dependence of the model is an essential ingredient is estimating the EMF $\Phi$ (\S\ref{section:EMF}). There is no photospheric dynamo. Although the physics is quite different, the numerical values of $I$ and $\Phi$ are similar in the two models. These parameters are constrained by the requirement that power, $I\Phi$, be equal to that released in a flare, and that the current be of order that inferred from magnetograms, $I=10^{11}$--$10^{13}\,$A.

A more subtle difference concerns the dissipation. In circuit models the energy release it attributed to Ohmic dissipation, described by a resistance in the circuit. The quadrupolar model is intrinsically time dependent, and the conventional circuit equation for energy does not apply to a time-dependent model. One can use energy arguments only by comparing the initial and final states, which are assumed time-independent. The EMF, $\Phi$, is (minus) the rate of change of the magnetic flux, and the counterpart of a resistance is played by (minus) the time derivative of the effective inductance, which decreases as the effective current path length decreases.

\section{Flare-associated EMF}
\label{section:EMF}

The EMF that drives a flare must be associated with the changing magnetic field.  In this section, three idealized models for a time-varying magnetic field are used to estimate the EMF. 

\subsection{EMF associated with magnetic energy release}

The first estimate of $\Phi$ is based on the integrated form  of $\curl{\bf E}=-\partial{\bf B}/\partial t$ applied to the time-varying magnetic field within the energy-release region. 

Consider a model in which the time-varying magnetic field is confined to a box with sides of length $L_x,L_y,L_z$, that is, the region $-L_x/2<x<L_x/2$, $-L_y/2<y<L_y/2$, $-L_z/2<z<L_z/2$. Let the magnetic field be  along the $z$ axis with 
\be
B_z(x,t)=B_0(2x/L_x)f(t),
\label{Bz}
\ee 
where $f(t)$ is a decreasing function over a characteristic time $T$, e.g., $f(t)=\exp(-t/T)$ or $1-t/T$. The inductive electric field is along the $y$ axis, with
\be
E_y(x,t)-E_y(0,t)={\dot f}(t)B_0(x^2/L_x),
\label{Ey}
\ee 
where the dot denotes the time derivative. The potential, $\Phi(x,t)=-{\dot f}(t)B_0(x^2L_y/L_x)$, across the box along the $y$~axis, is symmetric on either sides of the plane $x=0$. The current density (obtained by integrating the induction equation ignoring the displacement current) is along the $y$-axis with $J_y(t)=-B_0(2/\mu_0L_x)f(t)$, implying a current $I(t)=-2B_0L_zf(t)/\mu_0$ independent of $x$. The magnetic flux has one sign for $x>0$ and the opposite sign for $x<0$, with $\Psi_B=\pm B_0L_xL_yf(t)/2$ in the two halves. The rate of change of the magnetic energy, $dW_B/dt$, is given either by the time derivative of $W_B$, obtained by integrating the magnetic energy density, $B_z^2(x,t)/2\mu_0$, over the box, or by averaging $I(t)\Phi(x,t)$ over the box, either of which gives $dW_B/dt={\dot f}(t)f(t)B_0^2L_xL_yL_z/6\mu_0$. 

This model implies an EMF of order $\Phi=B_0L_xL_y/T$, where $T$ is the duration of the flare. Fiducial numbers are discussed below, and one choice is $B_0=10^{-2}\,$T, $L_x=L_y=10^7\,$m, $T=10^2\,$s, implying $\Phi=10^{10}\,$V. That such a large potential is needed to drive a flare has been known for many decades \citep{S33}.

The neglect of the displacement current in the foregoing calculation is well justified for $(L_x/cT)^2\ll1$. However, as discussed in paper~2, the displacement current drives a polarization current that is a factor $c^2/v_A^2$ larger, and the condition for the neglect of this current, $(L_x/v_AT)^2\ll1$, is not well justified. It is argued in paper~2 that the cross-field current associated with the energy release is the polarization current, and hence that the displacement current plays an essential (albeit indirect) role in the energy release.

\subsection{Time-varying inductance}

The second estimate of the EMF is based on the rate of change of the magnetic flux enclosed by the circuit around which the current flows. As argued above, the change in stored magnetic energy can be attribute to a reduction in the effective inductance, $L(t)=\mu_0\ell(t)$, of the circuit, requiring a reduction in the effective length, $\ell(t)$, of the current path. The magnetic flux in such a model is 
\be
\Psi_B(t)=L(t)I,
\label{PsiBt}
\ee
implying that the EMF is 
\be
\Phi(t)=-{\dot L}(t)I=-\mu_0{\dot\ell}(t)I.
\label{Phit}
\ee
One can write this in the form of Ohm's law, $\Phi(t)=\Sigma_0(t)I$, with an effective impedance $\Sigma_0(t)=\mu_0v_0(t)$, $v_0(t)=-{\dot\ell}(t)$. Further evaluation of this model requires an estimate of $\ell$, which depends on the geometry of the circuit.

Changes in the stored magnetic energy due to changes in the geometry are important in laboratory plasma experiments, and can be modeled through changes in self and mutual inductances \citep{BTS78}. Any distribution of currents in the corona can be approximated by a set of sub-currents, $I=I_1+I_2+\cdots$ say, such that the magnetic energy can be written as the sum $\half(L_1I_1^2+L_2I_2^2+\cdots)+M_{12}I_1I_2+\cdots$, where $L_1,L_2,\cdots$ are self inductances and $M_{12},\cdots$ are mutual inductances. This sum is written as $\half LI^2$ to define $L$. When the sub-currents are close together, the self and mutual inductances are nearly equal (to $L(t)$) and the mutual inductances decrease with increasing separation between the sub-currents. Assuming that a single value of $L(t)$ suffices, simple models for the geometry \citep{BTS78,M97} imply that $\ell$ differs from the actual length of the current path by a factor of order unity, which can be ignored for semi-quantitative purposes. One can then estimate $v_0=\Delta\ell/T$, where $\Delta\ell$ is the difference between the initial and final lengths of the current path. Thus this estimate gives $\Phi=\mu_0\Delta\ell I/T$. The fiducial numbers discussed below suggest $\Delta\ell=10^7\,$m, $I=10^{11}\,$A, $T=10^2\,$s, implying $\Phi=10^{10}\,$V, consistent with the first estimate.

\subsection{Propagating flux tube}

The third estimate of $\Phi$ is based on transfer of magnetic flux and associated current from the initial to the final flux tubes. Reconnection of force-free flux tubes in 3D leads to a portion of the initial flux tube propagating from its initial location to its final location, as illustrated in the quadrupolar model, Fig.~\ref{fig:quadrupolar}b. The effective length, $\ell(t)$, of the flux tube reduces, at the rate $v_0(t)$, as it propagates. This model gives an alternative way of estimating $\Phi(t)=-{\dot L}(t)I=-\mu_0v_0(t)I$ by providing a geometric model for $v_0(t)$.

Consider either of the triangles in Fig.~\ref{fig:quadrupolar}b, which are projections onto the solar surface. This suggests approximating the reconnecting flux tube as a triangle. Assuming that the final flux tube is in a vertical plane, whose base is the dashed line, the initial flux tube has the same base, and is in azimuthal plane, at an angle $\Delta\phi$ say to the vertical. Let the base be of length $\ell_0$, and the initial and final triangles being equilateral with sides $\ell_1/2,\ell_2/2$, respectively, with perpendicular bisectors of length $d_1,d_2$, respectively. A flux tube in the process of being transferred is at $0<\phi<\Delta\phi$ with sides $\ell/2$ and perpendicular bisector $d$. It is convenient to introduce the angle $\zeta$, such that $\ell=2d\sin\zeta$. One has ${\dot\ell}=2{\dot d}\sin\zeta$. The velocity of the apex of the triangle may be identified as $v_0$; it has contributions from the rate of change of both $\zeta$ and $\phi$, although only the former contributes to $\dot\ell$. To within a factor of order unity, one may estimate $v_0\approx{\dot d}\approx{\dot\ell}$. With ${\dot\ell}=-\Delta\ell/T$, this model reproduces the results of the other two models.

\subsection{Fiducial numbers for a flare}

As fiducial values, consider a moderately powerful flare with an energy $I\Phi=10^{21}\,$W transferred to energetic electrons over a time  $T=10^2\,$s, due to a current  $10^{11}\,$A and an EMF $\Phi=10^{10}\,$V. These numbers are consistent with a magnetic field $B=10^{-2}\,$T in a box with sides $\ell=10^7\,$m, with the reduction in the effective length, $\Delta\ell$, of the current path being of order $10^7\,$m. Assuming the implied effective impedance, $10^{-1}\rm\,\Omega$, to be the Alfv\'enic impedance implies $v_A=10^5\rm\,m\,s^{-1}$ in the energy-release region. These numbers are also consistent with the stored energy being identified as $\half LI^2=10^{23}\,$J, with an inductance $L=\mu_0\ell=10\,$H. The actual numbers for a given flare need to be greater or smaller than these values to account for less and more powerful flares.

It should be emphasized that the high value of $\Phi$, of order $10^{10}\,$V, is essential, but that there is no direct evidence for it. This contrasts with a substorm, where the maximum energy of auroral electrons is plausibly determined by $e\Phi$. In flares  electrons are accelerated  to 10-$20\,$keV, which is $e\Phi/M$ with $M\approx10^6$. The factor is referred to here as the multiplicity, $M$, in the sense that the rate, ${\dot N}$, electrons are accelerated is $MI/e$. The multiplicity quantifies the long-standing ``number problem.''

\section{Application to substorms}
\label{section:substorms}

Application of the generic model to substorms \citep{A64,KR95} is discussed briefly here for background and comparison purposes.

\subsection{Description of a substorm}

The magnetic energy released in a substorm is stored in the Earth's magnetotail. Prior to a substorm, magnetic flux builds up in the magnetotail  due to reconnection on the  sunward side of the magnetosphere at a time when the direction of the magnetic field in the solar wind is opposite to the Earth's magnetic field there. Reconnection occurs at a rate determined by a coupling function \citep{G90,Letal04}. The solar wind drags the reconnected flux tubes around to the anti-sunward side, forming the magnetotail. In a steady state, there is slow reconnection in the magnetotail, with the magnetic flux being returned to the sunward side through the inner part of the magnetosphere. The magnetic energy stored in the magnetotail is associated with a cross-tail current that flows along the dawn-dusk line in the current sheet that separates the oppositely directed field lines above and below the sheet. This current closes (around nearly semi-circular paths) on the surface of the magnetotail above and below the current sheet. When the orientation of the magnetic field in the solar wind reverses (a ``southward turning''), the supply of magnetic flux to the tail is shut off, and the steady-state balance of stresses in the magnetotail can no longer be sustained. This is one plausible cause for the onset of explosive energy release in a substorm. The magnetic energy release is associated with magnetic reconnection or current disruption in the magnetotail \citep{L11}. The tension in the newly formed flux tube accelerates plasma both toward and away from the Earth. The observed earthward high speed flow correlates with the onset of a substorm \citep{BPL90,Petal10}. Such flow is a possible magnetospheric counterpart of mass flows, including jets and CMEs, associated with flares.

During the substorm, magnetic reconnection leads to partial ``dipolarization'' of the magnetic field, in the sense that some open field lines become closed, reducing the magnetic flux in the open field lines. The cross-tail current is partly redirected into a field-aligned current flowing out of the energy-release region, forming the so-called current wedge, cf. Fig.~\ref{fig:generator}. The up and down currents from opposite ends of the cross-tail current flow along converging flux tubes, closing across field lines in the ionosphere. There are actually two current wedges, one in each hemisphere, so that there are up and down currents above the current-closure regions in both hemispheres. The magnetic energy released is transported Alfv\'enically along field lines to the acceleration region, which is located far above the ionosphere in the upward current region. An analogous Alfv\'enic energy flow is assumed to occur in flares, but the location of the acceleration region in a flare is not as well determined.

In the acceleration region, energy is transferred to auroral electrons by a parallel electric field, which is highly structured, involving a large number of localized regions (over several Debye lengths), propagating regions with $E_\parallel\ne0$ \citep{Eetal98a,MK98}, modeled as phase-space holes \citep{Eetal98b} or as kinetic Alfv\'en waves \citep{Cetal03}. The acceleration can be so efficient that charge starvation occurs \citep{Eetal04}: the current-carrying field lines can become devoid of thermal plasma, forming an auroral cavity. This tends to produce a flux of precipitating (``inverted-V'') electrons with energy $\approx e\Phi$ and precipitation rate $dN/dt\approx I/e$. There are nearby associated downward current regions, in both hemispheres, where ions are accelerated downward and the electron acceleration is associated with intense electrostatic turbulence \citep{Eetal98a,Eetal03}. A major difference between flares and substorns is that the electrons in a flare are accelerated only to a small fraction of $e\Phi$. Why this is the case is a long-standing problem. A possible explanation is suggested in paper~2.

\subsection{Fiducial numbers for a substorm}

Substorms have an onset phase of a few minutes and a total duration of a few hours, implying $T$ in the range $10^2$--$10^4\,$s: $T=10^3\,$s is adopted here. The parallel (Birkeland) currents are about $10^5\,$A in quiet times, increasing to $>10^6\,$A during a substorm. The EMF must be greater that the maximum energy of the auroral electrons. In inverted-V events, the energy spectrum has a maximum of several keV. This suggests fiducial values $I=3\times10^5\,$A, $\Phi=10^5\,$V. The magnetic flux enclosed by the circuit around which the current flows is estimated as $\Delta\Psi_M=\Phi T=10^8\,$Wb. (The portions of the circuit around which the Birkeland currents flow do not contribute significantly to the changing $\Psi_M$.) The estimated number is consistent with a (near) magnetotail magnetic field of about $1\,$nT times that cross-sectional area of the magnetotail. 

Assuming that the internal impedance in the energy-release region is determined by $R_A$ there, the fiducial number $R_A=\Phi/I=0.3\rm\,\Omega$ implies an  Alfv\'en speed, $v_A=3\times10^5\rm\,m\,s^{-1}$. This is a relatively low value, which is possible for a region that is overdense (in a current sheet) with a weak magnetic field (around a magnetic null).

\section{Discussion}
\label{section:discussion}

The generic model for magnetic explosions presented in \S\ref{section:generic}  is based on a widely accepted model for magnetospheric substorms, and is adapted in \S\ref{section:current} and \S\ref{section:EMF}  to apply to solar flares. The  model has separate magnetic-energy-release and electron-acceleration regions, with Alfv\'enic energy transport between them. Such a model involves different physics on vastly different scales, and three scales are introduced: global, macro and micro. In this paper the emphasis is on the global scale, which necessarily involves integrated quantities and circuit-type concepts. An obvious advantage of a circuit model is that it allows semi-quantitative discussion of the energy storage and release in terms of currents, as opposed to appealing to release of an ill-defined free magnetic energy density. However, most existing circuit models do not reflect the essential physics assumed here: explosive release of magnetic energy stored in the corona. Although the model proposed is not a circuit model in the conventional sense: the current path is not fixed, there is no photospheric dynamo, and the energy release is not due to the resistive decay of a current. Rather, the magnetic field is changing as a function of time, and this causes an EMF that redirects pre-existing current around a new path. Energy release is attributed to a reduction in the stored magnetic energy through a reduction in the effective length of the current path. Reconnection is an essential ingredient in allowing the magnetic configuration to change, to one of lower energy, but it plays only a secondary role in the model.

One motivation for introducing a generic model for magnetic explosions  is to avoid what I regard as a misleading feature in the existing literature: the use of steady-state proxies to describe intrinsically time-dependent electrodynamic phenomena. The inductive electric field ($\curl{\bf E}\ne0$) and the displacement current are excluded by the steady-state assumption. When the time dependence is taken into account, these are nonzero and play important roles, which are not treated properly by the steady-state proxies. In this paper, this is shown to be the case for the EMF that drives the flare-associated current in a flare. Other roles that the inductive electric field and the displacement current play in flares are discussed in paper~2.

One purpose of a global model for a flare is to describe the energetics. The power released can necessarily be written as $I\Phi$, in terms of the current, $I$, and the EMF, $\Phi$. It is argued that appropriate fiducial values, for a flare with a power $10^{21}\,$W are $I=10^{11}\,$A and $\Phi=10^{10}\,$V. The ratio $\Phi/T$, which is attributed to a resistance in a conventional circuit model, is interpreted as an effective impedance $\mu_0v_0$, with $v_0=-{\dot\ell}$ due to the decreasing length of the current path. It is suggested that this impedance is of order the Alfv\'en impedance, $R_A=\mu_0v_A$. Fiducial number for a flare suggest $\Phi/I$ is of order $0.1\rm\,\Omega$, which may be interpreted as a relatively slow Alfv\'en speed, $v_A=10^5\rm\,m\,s^{-1}$, in the energy-release region, or perhaps as $v_0$ being significantly smaller than $v_A$ due to geometric factors.

\section*{Acknowledgments}

I thank Iver Cairns, Mike Wheatland and Mohammad Rafat for helpful comments.


\begin{thebibliography}{22}

\bibitem[\protect\citeauthoryear{Akasofu}{1964}]{A64}
Akasofu, S.-I. 1964, Planet Space Sci., 12, 273

\bibitem[\protect\citeauthoryear{Achwanden}{2004}]{A04}
Aschwanden, M. J.  2004, {\it Physics of the Solar Corona: An Introduction} Springer, Berlin.
\bibitem[\protect\citeauthoryear{Aschwanden \et}{1999}]{Aetal99}
{Aschwanden, M. J., Kosugi, T., Hanaoka, Y., Nishio, M., 
\& Melrose, D. B.}
{1999}
{ApJ {\bf 526}, 1026--1045}
\bibitem[\protect\citeauthoryear{Baumjohann, Paschmann \&  L\"uhr}{1990}]{BPL90}
Baumjohann, W., G. Paschmann, G., \& L\"uhr, H. 1990, 
J. Geophys. Res., 95, 3801
\bibitem[\protect\citeauthoryear{Benz \& G\"udel}{2010}]{BG10}
Benz, A. O,  \& G\"udel, M. 2010, ARAA, 48, 241
\bibitem[\protect\citeauthoryear{Book, Turchi \& Stein}{1978}]{BTS78}
Book, D. L., Turchi, P., \& Stein, D. L. 1979, J. Computational Phys. 33, 271
\bibitem[\protect\citeauthoryear{Brown \et}{2009}]{Betal09}
Brown, J. C., Turkmani, R., Kontar, E. P., MacKinnon, A. L., \& Vlahos, L. 2009,
A\&A, 508, 993
\bibitem[\protect\citeauthoryear{Chaston \et}{2003}]{Cetal03}
Chaston, C. C., Bonnell, J. W., Carlson, C. W., McFadden, J. P., Strangeway, R. J., \& Ergun, R. E. 2003 Geophys. Res. Lett. 30, 1289

\bibitem[\protect\citeauthoryear{Colgate}{1978}]{C78}
Colgate, 1978 ApJ 221, 1068
\bibitem[\protect\citeauthoryear{Dahlburg, Antiochos \& Norton}{1997}]{DAN97}
Dahlburg, R. B., Antiochos, S. K., \& Norton, D. 1997 
Phys. Rev. E 56, 2094
\bibitem[\protect\citeauthoryear{Emslie \& Sturrock}{1982}]{ES82}
Emslie, A. G., \& Sturrock, P. A. 1982, Sol. Phys. 80, 99
\bibitem[\protect\citeauthoryear{Ergun \et}{1998a}]{Eetal98a}
Ergun, R. E., {\it et al}.,1998a, Geophys. Res. Lett. 25, 2025
\bibitem[\protect\citeauthoryear{Ergun \et}{1998b}]{Eetal98b}
Ergun, R. E., {\it et al}., 1998b, Phys. Rev. Lett. 81, 826
\bibitem[\protect\citeauthoryear{Ergun \et}{2003}]{Eetal03}
Ergun, R. E.,  Andersson, L.,  Carlson, C.W., Newman, D. L. \&  Goldman, M. V. 2003
Nonlinear Processes in Geophysics  10, 45
\bibitem[\protect\citeauthoryear{Ergun \et}{2004}]{Eetal04}
Ergun, R. E., Andersson, L., Main, D.,  Su,  Y.-J., Newman, D. L., 
Goldman,  M. V., Carlson, C. W., Hull, A. J., McFadden, J. P., \&  Mozer, F. S. 2004
J. Geophys. Res., 109, A12220

\bibitem[\protect\citeauthoryear{Gonzalez}{1990}]{G90}
Gonzalez, W. D. 1990, Planet. Space Sci. 38, 627

\bibitem[\protect\citeauthoryear{Fletcher \& Hudson}{2008}]{FH08}
Fletcher, L. \& Hudson, H. S. 2008
ApJ 675, 1645.
\bibitem[\protect\citeauthoryear{Forbes \& Malherebe}{1986}]{FM86}
Forbes, T. G. , \& Malherbe, J. M. 1986, ApJ, 302, L67
\bibitem[\protect\citeauthoryear{Hardy, Melrose \& Hudson}{1998}]{HMH98}
{Hardy, S. J., Melrose, D. B., \& Hudson, H. S.}
{1998}
Publ. Astron. Soc. Aust. 15, 318
\bibitem[\protect\citeauthoryear{Haerendel}{1994}]{H94}
Haerendel, G. 1994, ApJS, 90, 765

\bibitem[\protect\citeauthoryear{Haerendel}{2007}]{H07}
Haerendel, G. 2007, J. Geophys. Res., 112, A09214
\bibitem[\protect\citeauthoryear{Holman {\it et al}.}{2011}]{Hetal11}
Holman, G.,  {\it et al}. 2011, SSR 159, 107
\bibitem[\protect\citeauthoryear{Kivelson \& Russell}{1995}]{KR95}
Kivelson, M.G., \& Russell, C.T. (eds.) 1995, Introduction to Space Physics, Cambridge University
Press.
\bibitem[\protect\citeauthoryear{Kontar, Hannah \& Bian}{2011}]{KHB11}
Kontar, E. P.,  Hannah, I. G.,  \& Bian, N. H. {2011} ApJ, 730, L22
\bibitem[\protect\citeauthoryear{Kontar {\it et al}.}{2011}]{Ketal11}
Kontar, E.,  {\it et al}. 2011, SSR 159, 301
\bibitem[\protect\citeauthoryear{Leka \et}{1996}]{Letal96}
Leka, K. D., Canfield, R. C., McClymont, A. N., \& van Driel-Gesztelyi, L. 1996,
ApJ 462, 547 
\bibitem[\protect\citeauthoryear{Linton}{ 2006}]{L06}
Linton, M. G. 2006, JGR 111, A12S09
\bibitem[\protect\citeauthoryear{Linton \& Longcope}{ 2006}]{LL06}
Linton, M. G., \& Longcope, D. W. 2006, ApJ 642, 1177
\bibitem[\protect\citeauthoryear{Longcope,  Guidoni \& Linton}{ 2009}]{LGL09}
Longcope, D. W., Guidoni, S. E., \& Linton, M. G. 2009
ApJ 690,  L18
\bibitem[\protect\citeauthoryear{Lopez \et}{2004}]{Letal04} 
Lopez, R. E., Wiltberger, M., Hernandez, S., \& Lyon, J. G. 2004 Geophys. Res. Lett 31, L08804

\bibitem[\protect\citeauthoryear{Lui}{2011}]{L11} 
Lui, A. T. Y. 2011 J. Geophys. Res. 116, A04214
\bibitem[\protect\citeauthoryear{Lyutikov}{ 2006}]{Ly06}
Lyutikov, M. 2006, MNRAS, 367, 1594
\bibitem[\protect\citeauthoryear{McClymont \& Fisher}{1989}]{MF89}
{McClymont, A. N., \& Fisher, G. H.}
{1989}
{Solar System Plasma Physics: Geophysical Monograph 54. Edited by J. H., Jr. Waite, J. L. Burch \& R. L. Moore. ISBN 0-87590-074-7; QC809.P5S65 1989. Published by the American Geophysical Union, Washington, DC USA, 1989, pp.\ 219--225}
\bibitem[\protect\citeauthoryear{MacKinnon \& Brown}{1989}]{MB89}
MacKinnon, A. L. \& Brown, J. C. 1989, Sol. Phys., 122, 303
\bibitem[\protect\citeauthoryear{McPherron \et}{1973}]{McPetal73}
McPherron, R. L., Russell, C. T., \& Aubry, M. P. 1973,
J. Geophys. Res., 78, 3131.
\bibitem[\protect\citeauthoryear{Melrose}{1995}]{M95}
{Melrose, D. B.}
{1995}
ApJ 451, 391
\bibitem[\protect\citeauthoryear{Melrose}{1997}]{M97}
{Melrose, D. B.}
{1997}
ApJ 486, 521
\bibitem[\protect\citeauthoryear{Melrose}{2009}]{M09}
Melrose, D. B.
{2009},
{in  R.A. Meyers (ed.)  {\it Encyclopedia of complexity and systems science}, Part~1 Springer, p.~21}
\bibitem[\protect\citeauthoryear{Melrose \& Khan}{1989}]{MK89}
Melrose, D. B., \& Khan, J. I. {1989}
A\&A {219}, 308.
\bibitem[\protect\citeauthoryear{Moreton \& Severny}{1968}]{MS68}
Moreton, G. E.,  \& Severny, A. B. 1968 Sol. Phys. 3, 282
\bibitem[\protect\citeauthoryear{Mozer \& Kletzing}{1998}]{MK98}
Mozer, F. S., \& Kletzing, C. A. 1998
Geophys. Res. Lett. 25, 1629
\bibitem[\protect\citeauthoryear{Nishio \et}{1997}]{Netal97}
Nishio, M., Yaji, K., Kosugi, T., Nakajima, H., \& Sakurai, T. 1997, ApJ 489, 976
\bibitem[\protect\citeauthoryear{Pu \et}{2010}]{Petal10}
Pu, Z. Y., \et 2010 J. Geophys. Res. 115, A02212
\bibitem[\protect\citeauthoryear{R\"onnmark}{2002}]{R02}
R\"onnmark, K. 2002, JGR 107, 1430
\bibitem[\protect\citeauthoryear{Sato}{1978}]{S78}
Sato, T. 1978, 
J. Geophys. Res., 83, 1042
\bibitem[\protect\citeauthoryear{Song \& Lysak}{2006}]{SL06}
Song, Y., \& Lysak, R. L. 2006, PRL 96, 145002
\bibitem[\protect\citeauthoryear{Spicer}{1982}]{S82}
Spicer, 1982 SSR 31, 351


\bibitem[\protect\citeauthoryear{Swann}{1933}]{S33}
Swann, W. F. G. 1933 Phys. Rev 43, 217

\bibitem[\protect\citeauthoryear{T\"or\"ok \et}{2011}]{Tetal11}
T\"or\"ok, T. , Chandra, R., Pariat, E., DŽmoulin, P., Schmieder, B., Aulanier, G., Linton, M. G., \& Mandrini, C. H. 2011
ApJ 728, 65


\bibitem[\protect\citeauthoryear{Treumann \& Baumjohann}{1997}]{TB97}
Treumann, R. A., \& Baumjohann, W. 1997 {\it Advanced Space Plasma Physics}, Imperial College
Press, London.

\bibitem[\protect\citeauthoryear{Uchida}{1996}]{U96}
Uchida, Y. 1996 Adv. Space Res., 17, 19

\bibitem[\protect\citeauthoryear{Uchida \et}{1999}]{Uetal99}
Uchida, Y., Hirose, S. Cable, S., Morita, S., Torii, M., Uemura, S., \& Yamaguchi, T. 1999,
Pub. Astron. Soc. Japan 51, 553.

\bibitem[\protect\citeauthoryear{Uralov}{1996}]{Ur96}
Uralov, A. M. 1996, Solar Phys. 168, 311


\bibitem[\protect\citeauthoryear{Wang}{2011}]{Wetal11}
Wang, S.,  Liu, C.,  Liu, R.,  Deng, N.,   Liu, Y. \&  Wang, H. 2011
``First Flare-related Rapid Change of Photospheric Magnetic Field Observed by Solar Dynamics Observatory''
arXiv:1103.0027


\bibitem[\protect\citeauthoryear{Wheatland \& Melrose}{1994}]{WM94}
Wheatland, M.S., \& Melrose, D.B.
{1994}
ApJ {47}, 361

\bibitem[\protect\citeauthoryear{White \et}{2011}]{Whetal11}
White, S. M., \et 2011, SSR 159, 225

\bibitem[\protect\citeauthoryear{Zharkova \et}{2011}]{Zetal11}
Zharkova, V. V., \et 2011, SSR159, 357

\end{thebibliography}
\end{document}